\documentclass[ALICE,manyauthors]{cernphprep}
\usepackage[comma,square,numbers,sort&compress]{natbib}
\usepackage{lineno}
\usepackage{xspace}
\usepackage{xcolor}
\usepackage{euscript}
\usepackage{graphicx}
\usepackage{dcolumn}
\usepackage{bm}
\usepackage{graphics}
\usepackage{hyperref} 
\usepackage{siunitx}
\usepackage[T1]{fontenc}
\usepackage{orcidlink}

\DeclareSIUnit\clight{\text{\ensuremath{c}}}
\DeclareSIUnit\eVc{\eV/\clight} 
\DeclareSIUnit\keVc{\keV/\clight} 
\DeclareSIUnit\MeVc{\MeV/\clight} 
\DeclareSIUnit\MeVcc{\MeV/\clight\squared} 
\DeclareSIUnit\GeVc{\GeV/\clight} 
\DeclareSIUnit\GeVcc{\GeV/\clight\squared} 
\usepackage[Symbol]{upgreek}

\usepackage[utf8]{inputenc}
\newcommand{\q}[1]{``#1''}

\begin{document}
%

\newcommand{\pp}           {pp\xspace}
\newcommand{\pbpb}           {Pb--Pb\xspace}

\newcommand{\roots}        {\ensuremath{\sqrt{s}}\xspace}
\newcommand{\pt}           {\ensuremath{p_{\rm T}}\xspace}
\newcommand{\px}           {\ensuremath{p_{\rm x}}\xspace}
\newcommand{\py}           {\ensuremath{p_{\rm y}}\xspace}
\newcommand{\etarange}[1]  {\mbox{$\left| \eta \right| < #1$}}
\newcommand{\yrange}[1]    {\mbox{$\left| y \right| < #1$}}
\newcommand{\dndy}         {\ensuremath{\mathrm{d}N_\mathrm{ch}/\mathrm{d}y}\xspace}
\newcommand{\dndeta}       {\ensuremath{\mathrm{d}N_\mathrm{ch}/\mathrm{d}\eta}\xspace}
\newcommand{\avdndeta}     {\ensuremath{\langle\dndeta\rangle}\xspace}
\newcommand{\dNdy}         {\ensuremath{\mathrm{d}N_\mathrm{ch}/\mathrm{d}y}\xspace}
\newcommand{\dEdx}         {\ensuremath{\textrm{d}E/\textrm{d}x}\xspace}
\newcommand{\kstar}        {\ensuremath{k^*}\xspace}
\newcommand{\mt}           {\ensuremath{m_{\rm{T}}}\xspace}
\newcommand{\St}           {\ensuremath{S_{\rm{T}}}\xspace}

\newcommand{\seven}        {$\sqrt{s}~=~7$~Te\kern-.1emV\xspace}
\newcommand{\thirteen}        {$\sqrt{s}~=~13$~Te\kern-.1emV\xspace}
\newcommand{\tev}          {Te\kern-.1emV\xspace}
\newcommand{\lumi}         {\ensuremath{\mathcal{L}}\xspace}

\newcommand{\ITS}          {\rm{ITS}\xspace}
\newcommand{\TOF}          {\rm{TOF}\xspace}
\newcommand{\ZDC}          {\rm{ZDC}\xspace}
\newcommand{\ZDCs}         {\rm{ZDCs}\xspace}
\newcommand{\ZNA}          {\rm{ZNA}\xspace}
\newcommand{\ZNC}          {\rm{ZNC}\xspace}
\newcommand{\SPD}          {\rm{SPD}\xspace}
\newcommand{\SDD}          {\rm{SDD}\xspace}
\newcommand{\SSD}          {\rm{SSD}\xspace}
\newcommand{\TPC}          {\rm{TPC}\xspace}
\newcommand{\TRD}          {\rm{TRD}\xspace}
\newcommand{\VZERO}        {\rm{V0}\xspace}
\newcommand{\VZEROA}       {\rm{V0A}\xspace}
\newcommand{\VZEROC}       {\rm{V0C}\xspace}
\newcommand{\Vdecay} 	   {\ensuremath{V^{0}}\xspace}

\newcommand{\ee}           {\ensuremath{\mathrm{e}^{+}\mathrm{e}^{-}}} 
\newcommand{\pip}          {\ensuremath{\pi^{+}}\xspace}
\newcommand{\pim}          {\ensuremath{\pi^{-}}\xspace}
\newcommand{\kap}          {\ensuremath{\mathrm{K}^{+}}\xspace}
\newcommand{\kam}          {\ensuremath{\mathrm{K}^{-}}\xspace}
\newcommand{\KK}           {\ensuremath{\mathrm{K}^{+}\mathrm{K}^{-}}\xspace}
\newcommand{\KKbar}           {\ensuremath{\mathrm{K}\overline{\mathrm{K}}}\xspace}
\newcommand{\phiPart}      {\ensuremath{\phi}\xspace}
\newcommand{\pbar}         {\ensuremath{\overline{\mathrm{p}}}\xspace}

\newcommand{\dbar}{$\overline{\mathrm{d}}$}
\newcommand{\kzero}        {\ensuremath{{\mathrm K}^{0}_{\mathrm{S}}}\xspace}
\newcommand{\lmb}          {\ensuremath{\Lambda}\xspace}
\newcommand{\almb}         {\ensuremath{\overline{\Lambda}}\xspace}
\newcommand{\Om}           {\ensuremath{\Omega^-}\xspace}
\newcommand{\Mo}           {\ensuremath{\overline{\Omega}^+}\xspace}
\newcommand{\X}            {\ensuremath{\Xi^-}\xspace}
\newcommand{\Ix}           {\ensuremath{\overline{\Xi}^+}\xspace}
\newcommand{\Xis}          {\ensuremath{\Xi^{\pm}}\xspace}
\newcommand{\Oms}          {\ensuremath{\Omega^{\pm}}\xspace}

\newcommand{\Led}         {Lednick\'y--Lyuboshits\xspace}
\newcommand{\Ledn}         {Lednick\'y--Lyuboshits approach\xspace}
\newcommand{\Nphi}         {\ensuremath{\mathrm{N}}\mbox{--}\ensuremath{\phi}\xspace}
\newcommand{\pphi}         {\ensuremath{\mathrm{p}}\mbox{--}\ensuremath{\phi}\xspace}
\newcommand{\apphi}        {\ensuremath{\overline{\mathrm{p}}}\mbox{--}\ensuremath{\phi}\xspace}
\newcommand{\pphiComb}     {\ensuremath{\mathrm{p}}\mbox{--}\ensuremath{\phi} \ensuremath{\oplus} \ensuremath{\overline{\mathrm{p}}}\mbox{--}\ensuremath{\phi}\xspace}

\newcommand{\pd}{\ensuremath{\mbox{p--d}}~}
\newcommand{\ApAd}         {\ensuremath{\mathrm{\overline{p}}\mbox{--}\mathrm{\overline{d}}}\xspace}
\newcommand{\ld}{\ensuremath{\mbox{$\Lambda$--d}}~}

\newcommand{\pKK}         {\ensuremath{\mathrm{p}}\mbox{--}(\ensuremath{\mathrm{K}^{+}\mathrm{K}^{-}})\xspace}

\newcommand{\pP}           {\ensuremath{\mathrm{p}\mbox{--}\mathrm{p}}\xspace}
\newcommand{\pL}           {\ensuremath{\mathrm{p}\mbox{--}\Lambda}\xspace}

\newcommand{\ApAP}         {\ensuremath{\mathrm{\overline{p}}\mbox{--}\mathrm{\overline{p}}}\xspace}

\newcommand{\pPComb}       {\ensuremath{\mathrm{p}\mbox{--}\mathrm{p} \oplus \mathrm{\overline{p}}\mbox{--}\mathrm{\overline{p}}}\xspace}

\newcommand{\pdComb}       {\ensuremath{\mathrm{p}\mbox{--}\mathrm{d} \oplus \mathrm{\overline{p}}\mbox{--}\mathrm{\overline{d}}}\xspace}

\newcommand{\pppair}           {\ensuremath{\mathrm{p}\mathrm{p}}\xspace}
\newcommand{\pdpair}           {\ensuremath{\mathrm{p}\mathrm{d}}\xspace}

\newcommand{\pppLair}           {\ensuremath{\mathrm{p}\mathrm{p}_{\Lambda}}\xspace}

\newcommand{\pLpair} {\ensuremath{\mathrm{p}\mbox{-}\Lambda}\xspace}

\newcommand{\Rpd} {\ensuremath{\mathrm{R}_\mathrm{pd}}\xspace}
\newcommand{\Rd} {\ensuremath{\mathrm{R}_\mathrm{d}}\xspace}
\newcommand{\Rp} {\ensuremath{\mathrm{R}_\mathrm{p}}\xspace}

\newcommand{\Singled} {\ensuremath{\mathrm{d}}\xspace}

\newcommand{\fReal} {\ensuremath{\Re(f_0) =\nolinebreak 0.85 \pm\nolinebreak 0.34\,(\mathrm{stat.}) \pm\nolinebreak 0.14\,(\mathrm{syst.})~\si{fm}}\xspace}
\newcommand{\fImag} {\ensuremath{\Im(f_0) =\nolinebreak 0.16 \pm\nolinebreak 0.10\,(\mathrm{stat.}) \pm\nolinebreak 0.09\,(\mathrm{syst.})~\si{fm}}\xspace}
\newcommand{\dZero} {\ensuremath{d_0 =\nolinebreak 7.85 \pm\nolinebreak 1.54\,(\mathrm{stat.})  \pm\nolinebreak 0.26\,(\mathrm{syst.})~\si{fm}}\xspace}

\newcommand{\NsigmaNoCorr} {\ensuremath{5.7 \pm\nolinebreak 0.8\,(\mathrm{stat.})  \pm\nolinebreak 0.5\,(\mathrm{syst.})~\sigma}\xspace }

\newcommand{\NsigmaNoCorrRange} {\ensuremath{4.7 - 6.6~\sigma}\xspace}

\newcommand{\VeffGauss} {\ensuremath{V_{\mathrm{eff}}=\nolinebreak 2.5\pm\nolinebreak 0.9 \,(\mathrm{stat.})\pm\nolinebreak 1.4\,(\mathrm{syst.})~\si{\MeV}}\xspace}
\newcommand{\muGauss}   {\ensuremath{\upmu=\nolinebreak0.14\pm\nolinebreak 0.06 \,(\mathrm{stat.})\pm\nolinebreak 0.09\,(\mathrm{syst.})~\si{fm^{-2}}}\xspace}

\newcommand{\alphaYuk}  {\ensuremath{\upalpha=\nolinebreak65.9 \pm\nolinebreak 38.0\,(\mathrm{stat.})\pm\nolinebreak 17.5\,(\mathrm{syst.})~\si{\MeV}}\xspace}
\newcommand{\AYuk}  {\ensuremath{A=\nolinebreak0.021\pm\nolinebreak 0.009\,(\mathrm{stat.})\pm\nolinebreak 0.006\,(\mathrm{syst.})}\xspace}
\newcommand{\gYuk}  {\ensuremath{g_{\Nphi} =\nolinebreak 0.14\pm\nolinebreak 0.03\,(\mathrm{stat.})\pm\nolinebreak 0.02\,(\mathrm{syst.})}\xspace}

\newcommand{\norm} {\ensuremath{\mathit{M} = 0.96}\xspace}

\newcommand{\realScatSigma}{\ensuremath{2.3\upsigma}\xspace}

\begin{titlepage}
\PHyear{2025}       
\PHnumber{075}      
\PHdate{27 March}  

\title{Accessing the deuteron source with pion--deuteron femtoscopy in Pb--Pb collisions at $\mathbf{\sqrt{s_{\textup{NN}}}\textup{ = 5.02}}$~TeV} 

\ShortTitle{Deuteron source with pion–deuteron femtoscopy}   

\Collaboration{ALICE Collaboration\thanks{See Appendix~\ref{app:collab} for the list of collaboration members}}
\ShortAuthor{ALICE Collaboration} 

\begin{abstract}
Femtoscopy of non-identical particle pairs has been instrumental for precision measurements of both two-particle sources and the final-state interactions in high-energy elementary and heavy-ion collisions. 
The majority of measurements assessing the source properties are based
on identical particle pairs, providing direct access to the characteristics of the single-particle source. 
The work in this paper demonstrates, via femtoscopy measurements of charged pion--deuteron pairs in Pb--Pb collisions at $\sqrt{s_{\rm NN}} = 5.02$~TeV, the feasibility of accessing
the characteristics of the single-particle femtoscopic source by using particle pairs with large mass differences such as pions and deuterons.  
The first experimental results of the measurement of deuteron source sizes in ultrarelativistic heavy-ion collisions are presented.  
The results show good agreement with the trend derived from other charged hadrons such as pions, kaons, and protons as a function of transverse mass, indicating similar source properties.
\end{abstract}
\end{titlepage}

\setcounter{page}{2} 
\section{\label{sec:level1}Introduction}

In ultrarelativistic heavy-ion collisions, various particle species emerge from the created fireball. 
Among them, non-composite light hadrons are produced predominantly in a thermal way during the hadronization phase at the chemical freeze-out, occurring at a system's temperature of approximately 150--160 MeV~\cite{Cabibbo:1975ig,Shuryak:1977ut,HotQCD:2014kol,HotQCD:2018pds,Borsanyi:2020fev,Grigoryan:2021org}.
The high temperature has sparked debates about whether
composite particles, such as (anti\nobreakdash-)deuterons, can form directly alongside hadrons during these thermal processes ~\cite{Mrowczynski:2016xqm,Sombun:2018yqh,Blum:2019suo,Bellini:2020cbj,Kachelriess:2020amp,Florkowski:2023uim}. 
This question arises because the production and stability of (anti-)deuterons in a hot thermal medium are unexpected, given their modest binding energy of only 2.2~MeV between the neutron and proton $-$ approximately 70 times smaller than the system temperature at chemical freeze-out ($\sim$156~MeV~\cite{Tch156}).
Instead, an alternative scenario suggests that (anti-)deuterons form through coalescence, where protons and neutrons, close in phase space in their final state, interact via the strong force during the hadronic phase, between chemical and kinetic freeze-out.
The production of light ions via the thermal or coalescence processes has been extensively studied experimentally via measurements of particle yield, spectra~\cite{pro1, pro2, pro3,ALICE:2021lsv}, elliptic flow~\cite{production6, production7, production77}, and fluctuations~\cite{production4444}. 
However, no single production model fully explains all observed trends, as different studies either lack sensitivity or support different mechanisms. Therefore, to gain a better understanding of this topic, further studies of both established and new observables are required.

Recently, there has been a notable interest in correlation studies involving (anti\nobreakdash-)deuterons~\cite{Mahlein:2023fmx,Mrowczynski:2020ugu, bhawani, PhysRevC.111.034903}. 
One method for such studies is the femtoscopy technique (see Refs.~\cite{Lednicky:2005af,Lisa:2005dd} and references therein), which is inspired by Hanbury Brown and Twiss interferometry~\cite{HanburyBrown:1956bqd} in astronomy used to measure the angular sizes of light sources. 
Femtoscopy is based on measurements of correlations as a function of the relative momenta between the two particles and enables the measurement of space--time characteristics of short-lived dynamic sources of the size of 
1--10~fm~\cite{Kopylov:1972qw,Kopylov:1973qq,Lednicky:1979ig} as well as the final-state interactions (FSI) between the two particles of the pair~\cite{STAR:2015kha,Acharya:2020asf, kppp, ALICE:2021szj}. 
The study of relative momentum correlations involving (anti\nobreakdash-)deuterons can therefore provide valuable insights into several open questions, including the interactions of nucleons with other hadrons~\cite{bhawani} as well as the production mechanism of composite particles in heavy-ion collisions which may affect the space--time properties of the source~\cite{Mrowczynski:2019yrr, ant:3, Mrowczynski:2020ugu}. 
Femtoscopic correlations can provide unique and complementary access to the properties of interactions in the low momentum regime ($<$ 200~MeV/$c$) to other techniques like scattering experiments or spectroscopy of exotic atoms~\cite{ALICE:2021szj, PhysRevD.110.032004}.

An important objective of femtoscopy is the measurement of the size of the homogeneity region of the particle-emitting source~\cite{Akkelin:1995gh}. 
This region is defined as the part of the source from which pairs of particles are emitted with similar velocity magnitude and direction, corresponding to the low relative momentum in the pair rest frame (PRF). 
The size of the homogeneity region follows two dependencies: it scales linearly with the cube root of charged-particle multiplicity density $\langle\mathrm{d}N_{\rm ch}/\mathrm{d}\eta\rangle^{1/3}$, where $\eta$ is the pseudorapidity, and exhibits a power-law scaling with the transverse mass $m_{\rm T}=\sqrt{p_{\rm T}^2+m_{0}^{2}}$, where $p_{\rm T}$ and $m_{0}$ are the transverse momentum and the rest mass of the particle, respectively~\cite{Adamova:2002wi,Adams:2004yc,Abelev:2009tp,Adam:2015vna,ALICE:2015hvw}.
The $m_{\rm T}$ scaling of the system size in heavy-ion collisions is a sign of the collectivity arising from the flow field of an expanding system
~\cite{Alice2011,alice2020} that is predicted by models based on relativistic hydrodynamics~\cite{yu2015}. 
The assumption of a common $m_{\rm T}$ scaling behavior for all particle species is particularly useful as one can estimate the expected source size for a given particle species for which experimental data is not available~\cite{Kisiel:2014upa,ALICE:2015hvw}. 
This assumption also greatly reduces uncertainties in the study of the interaction of the final-state particles that make up the pair. 
By constraining the source, one can use experimental data to fit phenomenological models and parameterize unknown or poorly constrained variables of the interaction (see Refs.~\cite{Acharya:2020asf,Fabbietti:2020bfg}). 
Additionally, it is possible to investigate the dynamics of the interaction itself, such as many-body scenarios~\cite{bhawani, pd_theory_3body_Viviani2023}.
However, it is not immediately evident whether composite objects such as light nuclei should follow the same $m_{\rm T}$ scaling trend as non-composite hadrons like pions, kaons, and protons~\cite{Mrowczynski:2019yrr}. 
In this study, we specifically test whether this scaling extends to deuterons, which, despite being composite objects, have a binding energy that is significantly lower than the characteristic system temperature. 
The validity of the $m_{\rm T}$ scaling assumption for deuterons is directly examined by extracting their source sizes from femtoscopic measurements of pion–deuteron pairs and comparing them to the expected trend for individual hadrons. 
Determining the deuteron source size is thus a central goal of this work, as it provides crucial input for understanding the space--time evolution of composite particles in heavy-ion collisions.

The structure of this paper is organized as follows. 
In Section 2, the experimental setup and data samples used in this analysis are described. 
Section 3 outlines the methodology for constructing the experimental correlation functions and details of the corrections applied. 
Section 4 provides a theoretical interpretation of the femtoscopic correlation functions, including a description of the models and formalisms employed. 
In Section 5, the results of the femtoscopic measurements and their implications are presented. 
Section 6 discusses the systematic and statistical uncertainties associated with the measurements. 
Finally, Section 7 summarizes the key findings of this study and highlights potential avenues for future research.

\section{\label{sec:DataSample} Data sample}
This study is based on the analysis of 337 million Pb--Pb collisions at a center-of-mass energy per nucleon--nucleon pair of \mbox{$\sqrt{s_{\rm NN}}=5.02$~TeV} recorded by ALICE (A Large Ion Collider Experiment)~\cite{ALICE:2008ngc} during the Run 2 data-taking period of the Large Hadron Collider (LHC)~\cite{Evans:2008zzb} in the year 2018. 
The collisions were selected by using a trigger system based on the two V0 detectors~\cite{Abbas:2013taa} composed of two arrays of 32 scintillator counters on either side of the nominal interaction point in the pseudorapidity ranges of $2.8 < \eta < 5.1$ (V0A), and $-3.7 < \eta < -1.7$ (V0C). 
A minimum-bias event at the center of the ALICE detector is recorded if, in coincidence to the crossing of the two LHC beams, hits in both V0 detectors are registered. 
Following the procedure described in Ref.~\cite{Abelev:2013qoq}, the amplitudes of signals measured in both V0 detectors are used to determine the centrality of the collision, expressed in percentages of the total inelastic hadronic cross section. 
The measurement presented here is performed in three intervals of event centrality: 0--10$\%$, 10--30$\%$, 30--50$\%$.
All three intervals consist of minimum-bias triggered events, while the central (0–10\%) and semi-central (30–50\%) intervals additionally include events from specific centrality triggers enhancing statistical significance~\cite{ALICEcentrality}.

In this work, the following central barrel sub-detectors of the ALICE apparatus are used to reconstruct tracks and to infer the corresponding particle identity: the Inner Tracking System (ITS~\cite{ALICE_collaboration_2010}), the Time Projection Chamber (TPC~\cite{ALME2010316}), and the Time Of Flight detector (\mbox{TOF~\cite{Akindinov:2013tea}}). 
All three of them are embedded in the homogeneous magnetic field of maximum 0.5~T provided by a large solenoid magnet.

The ITS consists of six cylindrical layers of silicon detectors placed closest to the beam pipe. 
Therefore, it is used for precise reconstruction of the primary vertex. To optimize detector performance, events with multiple reconstructed primary interaction vertices, known as pile-up events~\cite{pileup}, are excluded. The main detector of ALICE, responsible for tracking, momentum reconstruction of the tracks, and particle identification, is the TPC. 
The TPC is a large barrel gas detector filled with a 90$\%$--10$\%$ Ar-CO$_2$ gas mixture, placed around the ITS detector. 
It covers a radial area from approximately 85 to 247~cm from the beam pipe and provides an active volume of 88 m$^3$. 
The TPC acceptance covers the pseudorapidity range of $|\eta|$ < 0.9. 
The track reconstruction is based on detecting the radiation induced by the ionizing charged particle along its trajectory in up to 159 pad rows arranged radially and 18 sectors in azimuthal angle per TPC side. 
The particle identification (PID) is performed through the measurement of specific energy loss (d$E$/d$x$) of each track in the TPC. 
The difference between the measured and expected signals, assuming a certain particle identity is then calculated in units of the detector resolution, expressed as $N\sigma$. 
For this study, the momentum of the particles is determined based only on the track information coming from the TPC detector. 
Since the TPC’s PID is limited to low momenta, where the Bethe-Bloch curves are well separated, the TOF detector is additionally used for track identification in the momentum region exceeding 0.5~GeV$/c$ for pions and 1.3~GeV$/c$ for deuterons, respectively. 
The time-of-flight information is calculated as the difference between the arrival time registered in the TOF and the collision start time provided by the T0 detector~\cite{ALICE:2004ftm}. 
The time-of-flight of a given particle, together with its measured momentum and path length, can be used to infer its mass, whose distribution can then be associated with a specific particle species using $N\sigma$. 
It is a cylindrical shape detector built of Multigap Resistive Plate Chambers (MRPC) and it is located at $\sim$380 cm from the collision vertex in the radial direction. 
The TOF covers the pseudorapidity range of $|\eta|$ < 0.9.

In this analysis, in order to ensure uniform acceptance, all collisions are required to have a primary vertex, $V_{\rm{Z}}$, within $\pm$10~cm from the nominal interaction point along the beam ($Z$) axis and tracks are registered within a pseudorapidity range of $|\eta|<0.8$. 
The reconstruction of tracks is based on a minimum of 70 clusters deposited in the TPC detector. 
The transverse momentum $p_{\rm T}$ range of pions and (anti\nobreakdash-)deuterons, covers 0.1 $< p_{\rm T} <$ 1.5 GeV/$c$ and 0.8 $< p_{\rm T} <$ 2.0 GeV/$c$, respectively. 
Additionally, pions and antideuterons are accepted if their distance of closest approach (DCA) to the primary vertex is smaller than 2.4 cm and 3.2 cm in the direction transverse to the beam (XY) and along the beam (Z), respectively. 
A significant number of deuterons are produced by spallation processes in the detector material which typically do not point back to the primary collision vertex. 
In order to suppress this background, the DCA of deuterons is required to be smaller than $0.0105 +  0.0350 \times (p_{\text{T}}/(\rm{GeV}/\textit{c}))^{-1.1} ~\text{cm}$ in the XY plane and 1 cm in the Z direction. 
The track selection criteria of pion and (anti\nobreakdash-)deuteron are summarized in  Table~\ref{tab:trackcuts}. 

\begin{table}[tb!h]
\centering
\caption{Single track selection criteria for pions and (anti\nobreakdash-)deuterons.}
\label{tab:trackcuts}
\begin{tabular}{l|l|l|l}
  \hline
  \hline
  \multicolumn{4}{c}{Track selection} \\ \hline
   $\uppi^\pm$ $p_{\rm T}$  & \multicolumn{3}{c}{$0.1<p_{\rm T}<1.5$ GeV/$c$} \\ \hline
   d/$\overline{\rm d}$ $p_{\rm T}$  & \multicolumn{3}{c}{$0.8<p_{\rm T}<2.0$ GeV/$c$} \\ \hline
   $|\eta|$ & \multicolumn{3}{c}{$< 0.8$ }\\ \hline
  \hline
  \multicolumn{4}{c}{Pion ($\uppi^+$ and $\uppi^-$) selection} \\ \hline
    $N_{\sigma,\rm TPC}$ ($p < 0.5$~GeV/$c$) & \multicolumn{3}{c}{$< 3$} \\ \hline
    $\sqrt{N_{\rm \sigma, TOF}^{2} + N_{\rm \sigma, TPC}^{2}}$ 
    ($  p > 0.5$~GeV/$c$) 
    & \multicolumn{3}{c}{$< 3$}  \\ \hline
    $\rm DCA_{\rm Z}$ to primary vertex & \multicolumn{3}{c}{$ <3.2$ cm} \\ \hline
    $\rm DCA_{\rm XY}$ to primary vertex & \multicolumn{3}{c}{$ <2.4$~cm} \\
    \hline
  \multicolumn{4}{c}{Deuteron (d and $\rm \overline{d}$) selection} \\ \hline
    $N_{\sigma,\rm TPC}$ ($p < 1.3$~GeV/$c$) & \multicolumn{3}{c}{$< 2$} \\ \hline
    $N_{\rm \sigma, TOF}$,  $N_{\rm \sigma, TPC}$  ( $p >$ 1.3~GeV/$c$) & \multicolumn{3}{c}{$< 2$} \\ \hline
    $\rm DCA_{\rm Z}$ to primary vertex & \multicolumn{3}{c}{$< 1.0$ cm} \\ \hline
    $\rm DCA_{\rm XY}$ to primary vertex & \multicolumn{3}{c}{$<0.0105 +  0.0350 \times (p_{\text{T}}/(\rm{GeV}/\textit{c}))^{-1.1} ~\text{cm}$, $<$2.4~cm} \\ \hline  \hline

\end{tabular}
\end{table}

Moreover, tracks have to pass pair requirements to reduce the two-track effects, such as track merging, contributing to the measured distributions. 
For this purpose, similarly, as in Ref.~\cite{ALICE:2021szj}, a fraction of merged space points in the TPC are examined to determine whether the two tracks in the pair have a small relative proximity within the detector ($|\Delta \eta|<0.01$). 
Merged space points are defined by the distance between two tracks being less than 5 cm, and are evaluated at 1 cm intervals in the outward direction of the TPC (starting at 80~cm from the beam). 
Pairs with a merged fraction above 2\% are removed. 
To avoid spurious correlations coming from the misidentification of pairs originating from photon conversions, the invariant mass of the pair (assuming an electron mass hypothesis for both particles) is required to be larger than 0.002 GeV$/c^2$, and the relative polar angle, $\Delta\theta$, between the two tracks, should be larger than 0.008 rad.

\section{\label{sec:femtoscopy} Formalism of correlation functions}

The femtoscopy technique measures the correlation between two particles as a function of $k^*$, which is defined as the absolute value of half of the pair relative momentum, $k^* = (\vec{p_1^*}-\vec{p_2^*})/2$, where variable $p_1^*$ and $p_2^*$ are the momenta in the particle rest frame ($\vec{p_1}^*=-\vec{p_2}^*$). 
At low relative momenta, correlation functions become sensitive to the space--time separation of the particle emitters and to effects such as the (anti\nobreakdash-)symmetrization of the wave function (usually referred to as \q{quantum statistics}) and/or final state interactions (Coulomb and strong forces). 
The nature of the interaction as well as the particle source depend on the pair type under consideration, pair transverse mass, types of particles being collided, and the particle multiplicity of the collision~\cite{Lisa:2005dd,Lednicky:2005tb}. 
The femtoscopic correlation function can be expressed by the Koonin--Pratt formula~\cite{Koonin:1977fh,Pratt:1990zq} as 
\begin{equation}
C(\vec{k}^*) = \int S (\vec{r}^* )~\left | {\Psi  ( \vec{k}^{*},\vec{r}^{*} )} \right |^2 \mathrm{d}^3{r}^*,
\label{eq:cf_def}
\end{equation}
where $\vec{r}^*$ is the relative separation vector, $S (\vec{r}^*  )$ is
two-particle emitting source function, and $\Psi (\vec{k}^{*},\vec{r}^{*})$ is the pair wave function.

The source function $S(\vec{r}^*)$ of the homogeneity region describes the probability of emitting two particles at a relative distance $r^*$.  
It is typically represented as a three-dimensional spheroid with a Gaussian density profile parametrization~\cite{Broniowski:2008vp,Acharya:2017qtq,Acharya:2020nyr}. 
Such a source is then characterized with the so-called femtoscopic radii (width parameters) expressed either in a convoluted one-dimensional form, as is the case in this work, or directly in three dimensions~\cite{Alice2011}. 
The Gaussian source distribution is convoluted with the part representing the admixture of long-lived strongly decaying resonances (for pions e.g. $\omega$, $K^{0*}$~\cite{Kisiel:2006is}) present in the studied sample, which contribute to the non-Gaussianity of the experimentally observed source~\cite{Kisiel:2009eh}.
In our study, the statistical uncertainties of the data do not provide enough sensitivity for this effect and thus it is not considered in the fits. 
As a consequence, for the precision expected in the pion--deuteron measurement of this work, the usage of a Gaussian source, characterized by a single direction-averaged source size, is justified and sufficient.
According to hydrodynamical models~\cite{Huovinen:2006jp,Gale:2013da}, which 
mimic the evolution for the matter produced in ultrarelativistic heavy-ion collisions, the source is expected to be the same for same- and opposite-charge pairs, providing a single femtoscopic radius ($R_{\rm \uppi d}$) per centrality interval.

The pair wave function for non-identical particles accounts only for the effects of FSI. The pion--deuteron wave function is influenced by the Coulomb interaction, which dominates over the strong interaction, especially for same-charge pairs. 
As a result, the repulsive $\rm \uppi^{+}d$ interaction is approximated using a Coulomb-only solution. 
However, for opposite-charge pairs, the strong interaction plays a more significant role due to the attractive nature of the interaction. 
Therefore, given the well constrained scattering parameters~\cite{PhysRevC.58.R1869}, the strong interaction is included in the calculations for $\rm \uppi^{-}d$ pairs.
The presence of both Coulomb and strong forces in the interaction can be described with the Lednick\'y--Lyuboshitz formalism~\cite{Lednicky:1981su,Lednicky:2003mq,Lednicky:2009zza}, which approximate solutions of the interaction provide accurate descriptions for meson–nucleus systems in relativistic heavy-ion collisions. 
This formalism is based on the s-wave approximation of the asymptotic scattering wave function, which is fully described by the scattering parameters, specifically the scattering length ($f_0$) and the effective range ($d_0$). 
In this analysis, the zero-effective-range approximation ($d_0=0$) is used. The $\rm \uppi^{-}d$ scattering length of the strong interaction has been established as $f_0=-0.0382( \pm0.0007)+i\cdot0.0092( \pm0.0010)$~fm in precise x-ray spectroscopy of pionic deuterium experiments ~\cite{CHATELLARD1997855,PhysRevC.58.R1869} and agrees with theoretical expectations ~\cite{Meissner:2005bz,Baru2011PrecisionCO}. 
The pair wave functions for same- and opposite-charge pairs are defined in Appendix~\ref{sec:appendix:LL}. 


\section{\label{sec:DataAnalysis} Experimental correlation functions}
The correlation function is determined experimentally as the ratio of distributions of pairs as a function of $k^*$ constructed by using particles produced in the same event $A(k^*)$  and pairs originating from mixed-events $B(k^*)$~\cite{KOPYLOV1974,Acharya:2017qtq}. 
In the former case, pairs are physically correlated since both particles are produced in the same collision, whereas in the latter case, pairs do not exhibit any physical correlations. Both distributions contain effects related to the limited detector acceptance; therefore, the event mixing technique is used to account for them.
Ensuring similar event characteristics, the two particles forming pairs in the $B(k^*)$ distribution originate from events with a $V_{\rm{Z}}$ difference of less than 2~cm and centrality percentages differing by no more than 2.5$\%$ for central collisions (0–10$\%$) and 5$\%$ for semi-central collisions (10–30$\%$ and 30–50$\%$). 
To reduce statistical effects, particles from one event are mixed with those from up to 10 different similar events.
The experimental correlation function takes the form of $C_{\text{exp}}(k^*)=\EuScript{N}~A(k^*)/B(k^*)$, where $\EuScript{N}$ is a free normalization factor constrained in the  $0.15$--$0.35$~GeV$/c$ interval of $k^*$, where no femtoscopic signal is expected and non-femtoscopic effects are minimal. 
In this region, the correlation function is flat and approaches unity. 
The femtoscopic effects in the correlation function manifest as deviations from unity~\cite{Acharya:2020asf} which is particularly evident at low relative momentum. 
Consequently, the effects of the interaction and the femtoscopic source distributions are studied in the $k^* < 0.12~\text{GeV}/c$ region, referred to as low $k^*$ in this work.

The experimental correlation functions have been corrected for particle misidentification, secondary contamination from weak decays, e.g. $\rm K^0_S$ or $\rm \Lambda$, and deuterons that are produced through spallation processes in the detector material. The impact of these effects has been estimated using a data-driven approach and Monte Carlo (MC) simulations anchored to the detector performance during the corresponding data-taking period~\cite{ALICE:2021szj, PhysRevC.99.024001}. The particles originating from strongly decaying resonances form a long tail in the source distribution and have also been considered. The correction has been estimated based on a fraction of the Gaussian core in the full source, following the method described in Ref.~\cite{Kisiel:2009eh}. This method has already been used in the femtoscopic analysis e.g. pion--kaon and kaon--proton pairs by ALICE~\cite{Acharya:2020nyr,ALICE:2021szj}. 
The total correction ($corr$), is expressed as 
\begin{equation}
    corr = p_{\rm \pi}\cdot p_{\rm d}\cdot f_{\rm \pi}\cdot f_{\rm d}\cdot g_{\rm{\pi d}},
    \label{eq:corr2}
\end{equation}
where $p$ corresponds to the fraction of correctly identified particles, $f$ is the fraction of particles with primary origin, and $g$ is the source Gaussianity. The $corr$ factor has been calculated for each pair as a $k^*$-dependent parameter, which, on average, in the low $k^*$ range was found to be 61.6$\%$, 63.7$\%$, 65.1$\%$ for $\rm\uppi^{+}d$ and $\rm\uppi^{-}d$ pairs and 67.9$\%$, 66.3$\%$, 64.0$\%$ for $\rm\uppi^{+}\overline{d}$ and $\rm\uppi^{-}\overline{d}$ in centrality bins, 0--10\%, 10--30\%, 30--50\%, respectively. The correlation functions have also been corrected for a physical background ($baseline$) which arises from the collective expansion of the collision system. This background is estimated by extrapolating a first-order polynomial fit to the data points in the range 0.15$<k^*<$0.35 GeV/$c$, and equals $-0.008k^{*}+1.002$, $-0.017k^*+1.004$, and $-0.018k^*+1.004$ for 0--10\%, 10--30\%, and 30--50\%, respectively. The baseline is removed from the correlation function by division. The feed-down and residual correlations from particles that are not of primary interest are negligible and are assumed to follow flat distributions in pion-deuteron measurements. The genuine correlation function is then extracted as follows
\begin{equation}
    C(k^*) = \left(\frac{C_{\text{exp}}(k^*)-1}{corr(k^*)}+1\right)/ baseline(k^*).
    \label{eq:corr}
\end{equation}

\section{Results}
The presented analysis is performed for $\rm \uppi^{-}d$, $\rm\uppi^{+}\overline{d}$, $\rm\uppi^{+}d$, and $\rm\uppi^{-}\overline{d}$ pairs. Pairs and their charge conjugates were combined, as no systematic deviations between them were observed within uncertainties. The total number of pairs contributing to a low $k^*$ range in both cases of $\rm \uppi^{-}d$ $\oplus$ $\rm\uppi^{+}\overline{d}$, and $\rm\uppi^{+}d$ $\oplus$ $\rm\uppi^{-}\overline{d}$ amounts to $\num{3.6e7}$, $\num{3.7e6}$ and $\num{2.9e6}$ in $0-10\%$, $10-30\%$, and $30-50\%$ centrality intervals, respectively. 

\begin{figure*}[tb!]
    \centering
    \includegraphics[width=1.0\textwidth]{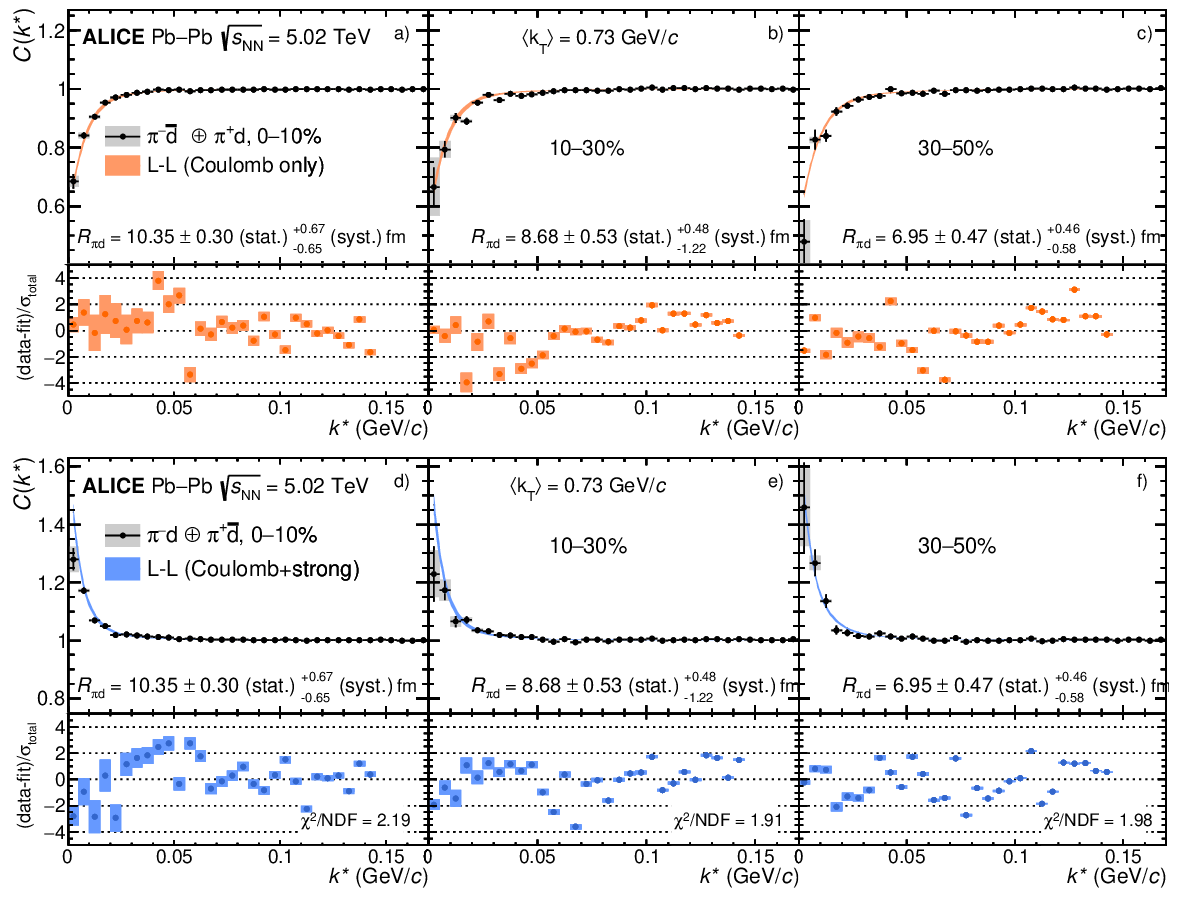}
    \caption{
    Upper row: $\rm \uppi^{+}d \oplus \uppi^{-}\overline{d}$ correlation functions in three centrality classes with Coulomb (Lednick\'y--Lyuboshitz, denoted as L-L) fits shown as orange bands. 
    Bottom row: $\rm \uppi^{-}d \oplus \uppi^{+}\overline{d}$ correlation functions in three centrality classes with L--L fits including both Coulomb and strong interaction shown as blue bands. 
    The data are shown after non-femtoscopic background subtraction and corrections (see text for more details). 
    The statistical and systematic uncertainties of the experimental data are represented by lines and boxes, respectively.
    Bandwidths represent the systematic uncertainties related to the fit. The bottom sub-panels display the data-to-fit differences normalized by the total uncertainty $\sigma_{\rm total}$ of the experimental data.
    }
    \label{fig:cfs}
\end{figure*}

Figure~\ref{fig:cfs} shows the experimental correlation functions for the same- and opposite-charge particle pairs, after baseline correction and background removal following Eq. (\ref{eq:corr}). The functions of $\rm \uppi^{-}\overline{d} \oplus \uppi^{+}d$ pairs exhibit an interaction that is repulsive in the low $k^{*}$ range where points are below unity. In that region, the $\rm \uppi^{-}d\oplus \uppi^{+}\overline{d}$ pairs show the opposite behavior of attractive interaction with points consistently above unity. 
The experimental data are further used to determine the femtoscopic source radii of pion-deuteron pairs for each centrality interval considered in this work. For this purpose, the employed fitting procedure is analogous to that described in Ref.~\cite{ALICE:2021szj}. The experimental correlation functions are compared with precomputed model functions, and the best source parameters are determined based on the lowest $\chi^2$/ndf within the range of $k^* < 0.12$ GeV/$c$ and are shared among same- and opposite-charge pairs of particles.
The precomputed theoretical correlation function is smeared using an experimental response function modeled with a Gaussian distribution, with its width determined from the low $k^*$ range of the resolution matrix and weighted by the number of pairs in the corresponding $k^*$ bin. 
The resolution matrix, representing the detector response, accounts for the finite momentum resolution of the ALICE detector, which affects the experimentally measured correlation function. 
It is derived from a Monte Carlo HIJING~\cite{Wang:1991hta} simulation coupled to GEANT4~\cite{Brun:1994aa} transport code, which provides a realistic description of the ALICE apparatus. 

The bands in Fig.~\ref{fig:cfs} show the result of the simultaneous fit to same- and opposite-charge particle pairs, described with the Lednick\'y--Lyuboshitz~\cite{Lednicky:1981su, Lednicky:2003mq} formalism, considering a Coulomb-only and a Coulomb– strong parametrization of the interaction, respectively. The models used for fitting provide a qualitatively good description of the experimental correlation functions and the fit bands follow closely the data points. The fit assumes that the source has the same size for same- and opposite-charge pairs at a given centrality. The two-particle $R_{\rm \uppi d}$ radii increase from \mbox{6.95 $\pm$ $0.47$(stat.) $^{+0.46}_{-0.58}$(syst.)~fm} for the 30--50$\%$ centrality, then to \mbox{8.68 $\pm$ $0.53$(stat.) $^{+0.48}_{-1.22}$(syst.)~fm} for the 10--30$\%$, and reaching the highest value of \mbox{10.35 $\pm$ $0.30$(stat.) $^{+0.67}_{-0.65}$(syst.)~fm} for the 0--10$\%$ centrality, respectively. The observed change of femtoscopic source sizes in different centrality intervals is expected as it has been observed already for femtoscopic measurements for other pair types in heavy-ion collisions~\cite{ALICE:2015hvw} and hydrodynamic models~\cite{Kisiel:2014upa,Kisiel:2018wie}. In Sec.~\ref{sec:syst_unc}, the details of the systematic uncertainty for the two-particle radii measurements are discussed. 

\begin{figure*}[tb!]
    \centering
    \includegraphics[width=0.8\textwidth]{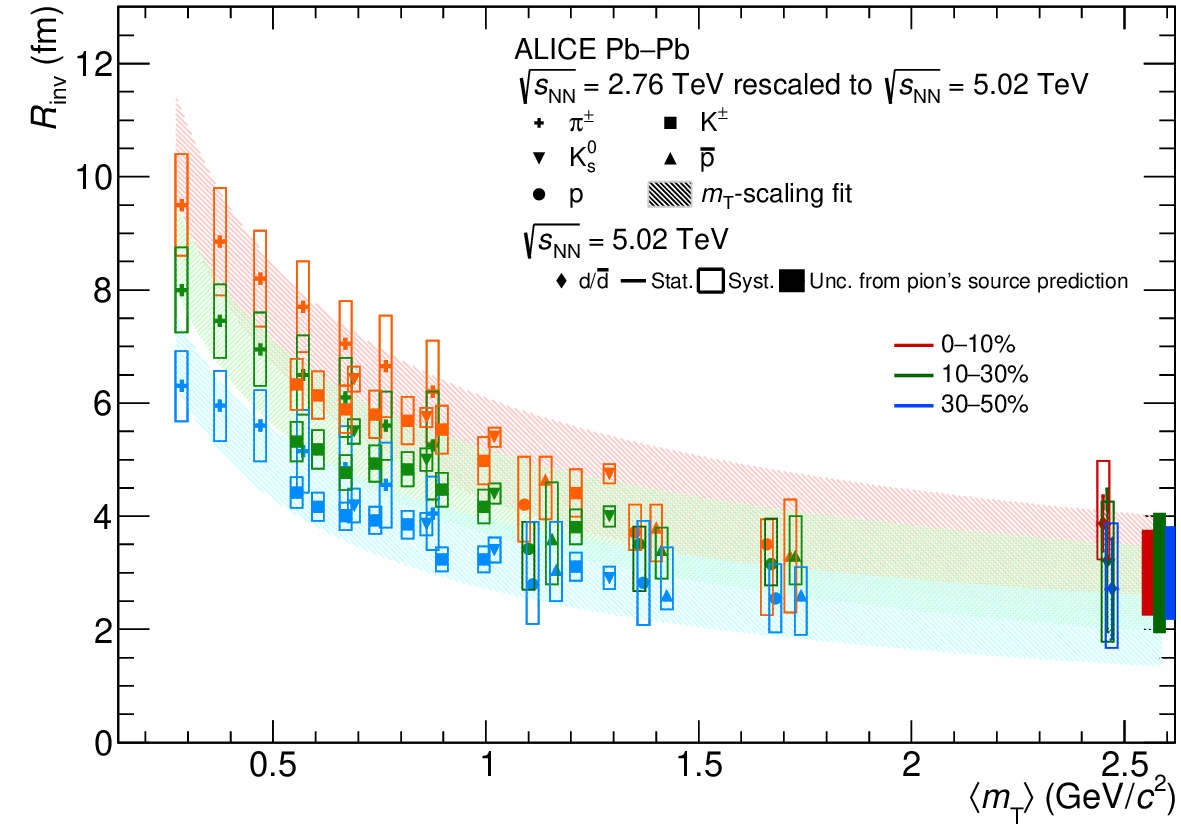}
    \caption{
    Values of $R_{\rm inv}$ as a function of $m_{\rm T}$ for Pb--Pb collisions at $\sqrt{s_{\text{NN}}}~=$~5.02 TeV for d and $\rm \bar{d}$,
    together with other mesons and baryons ($\uppi^{\pm}$, K$^{\pm}$, K$^{0}_{\rm S}$, p, $\rm \bar{p}$) taken from Ref.~\cite{ALICE:2015hvw} and rescaled to $\sqrt{s_{\text{NN}}}~=$~5.02 TeV.
    The red, green, and blue colors stand for $0-10\%$, $10-30\%$, and $30-50\%$ centrality intervals, respectively. The bands correspond to power-law fits of the femtoscopic sources for all previously measured particles, with the width representing the uncertainty in the pion source measurements.
    Statistical uncertainties are shown as lines. Systematic uncertainties arising from the analysis settings and from the pion source calculations based on rescaled data are shown with open and full boxes, respectively.
    }
    \label{fig:cfs2}
\end{figure*} 

The properties of particle emission are characteristic of different particle species. Consequently, the two-particle source parameter $R_{\rm \uppi d}$ is determined by the convolution of the corresponding single-particle radii~\cite{Kisiel:2009eh}, as described by the following formula.
\begin{equation}
R_{\uppi d} = \sqrt{R_{\uppi}^2+R_{d}^2},
\label{eq:rpi}
\end{equation}
where $R_{\rm \uppi}$ and $R_{\rm d}$ are the femtoscopic radii of particles of the pair that have similar velocities. When applied to a pair of identical particles, this equation can be simplifies to $\sqrt{2}R_{\rm inv}$, where $R_{\rm inv}$ represents the invariant single-particle source size for a given particle species.
Equation (\ref{eq:rpi}) can be further inverted to determine the source parameter $R_{\rm d}$. 
Here, the necessary inputs are the source radii of the pion--deuteron pair, $R_{\uppi \rm d}$, which is measured directly in this study, and the single-particle pion source radii, $R_{\uppi}$, obtained from the established $m_{\rm T}$-scaling~\cite{ALICE:2015hvw}. 
The latter was measured at $\sqrt{s_{\rm NN}} = 2.76$~TeV; therefore, to apply it in this work, the source radii are rescaled to a collision energy of $\sqrt{s_{\rm NN}} = 5.02$~TeV using the well-known linear scaling of femtoscopic sizes with the cube root of the charged-particle multiplicity density, given by
\begin{equation}
R_{\rm inv}=a\cdot\sqrt[3]{\langle\mathrm{d}N_{\rm ch}/\mathrm{d}\eta\rangle}+b,
\end{equation}
where $a$ and $b$ are free parameters derived for each particle species based on already measured femtoscopic sizes. The necessary scaling values of $\langle\mathrm{d}N_{\rm ch}/\mathrm{d}\eta\rangle^{1/3}$ are measured at both energies~\cite{PhysRevLett.106.032301, PhysRevLett.116.222302}. To get the $m_{\rm T}$ of pions in the non-identical pion--deuteron pairs sample, where differences in mass are large, only those particles that contribute to low $k^*$ are considered. These particles propagate at comparable velocities, contributing to the observed correlation effects. The average $m_{\rm T}$ of pions that build pairs with (anti\nobreakdash-)deuterons of this study is measured to be 0.24~GeV/$c^2$. Such $m_{\rm T}$ value access the ultra-soft pion limit and therefore the $m_{\rm T}$-scaling dependence of the rescaled pions is extrapolated with a linear fit function following the conclusion of Ref.~\cite{PhysRevC.110.034904}. 
The extrapolation of the $m_{\rm T}$-dependence of the pion femtoscopic source size is described by the function \mbox{$R_{\pi}(m_{\rm T}) = a\cdot m_{\rm T} + b$ fm}, where $a$ and $b$ are parameters that depend on centrality and are summarized in Tab.~\ref{tab-pion}. Consequently, for the measured pion $m_{\rm T}$ value of 0.24~GeV/$c^2$, the source radii of pions ($R_{\pi}$) are \mbox{9.60 $\pm$ 0.06 (stat.) $\pm$ 0.88 (syst.) fm}, \mbox{8.09 $\pm$ 0.05 (stat.) $\pm$ 0.70 (syst.) fm}, \mbox{6.45 $\pm$ 0.04 (stat.) $\pm$ 0.65 (syst.) fm} for 0--10$\%$, 10--30$\%$, 30--50$\%$ centrality, respectively.

In Fig.~\ref{fig:cfs2}, the femtoscopic single-particle source sizes of the (anti\nobreakdash-)deuteron are presented, together with the source sizes of other meson and baryon rescaled to the energy$\sqrt{s_{\text{NN}}}~=$~5.02 TeV~\cite{ALICE:2015hvw}. The $m_{\rm T}$ of (anti\nobreakdash-)deuterons has been measured to be 2.47~GeV/$c^2$, similar to the case of pions, by considering only the $\langle p_{\rm T}\rangle$ of particles that effectively contribute to low $k^*$. The bands in the plot correspond to power-law fits for all previously measured particles that are thermally produced. The band width corresponds to the uncertainty in the pion source radii. The values of $R_{\rm d}$ obtained in this study are equal to \mbox{$3.88\pm0.48$  (stat.) $^{+1.10}_{-0.64}$(syst.)~fm}, \mbox{$3.29\pm1.21$ (stat.) $^{+0.97}_{-1.52}$ (syst.)~fm}, \mbox{$2.72\pm0.99$ (stat.) $^{+1.15}_{-1.05}$ (syst.)~fm} for centralities 0--10$\%$, 10--30$\%$, 30--50$\%$, respectively.

\begin{table}[h]
\label{tab-pion}
\centering
\caption{Source radii of pions ($R_{\pi}$)  and deuterons ($R_{\rm d}$) for different centrality intervals.}
\begin{tabular}{|c|c|c||c|}
\hline
\textbf{Centrality} & \mbox{$R_{\pi}(m_{\rm T}) = a\cdot m_{\rm T} + b$} & $R_{\pi}$ (0.24~GeV/$c$) \textbf{(fm)} & \textbf{$R_{\rm d}$ (fm)} \\
\hline
0--10\% & $a=-5.62, b=10.94$ & $9.60 \pm 0.06\text{(stat.)} \pm 0.88\text{(syst.)}$ & $3.88 \pm 0.48\text{(stat.)} ^{+1.10}_{-0.64}\text{(syst.)}$ \\
\hline
10--30\% & $a=-4.70, b=9.23$ & $8.09 \pm 0.05\text{(stat.)} \pm 0.70\text{(syst.)}$ & $3.29 \pm 1.21\text{(stat.)} ^{+0.97}_{-1.52}\text{(syst.)}$ \\
\hline
30--50\% & $a=-3.76, b=7.36$ & $6.45 \pm 0.04\text{(stat.)} \pm 0.65\text{(syst.)}$ & $2.72 \pm 0.99\text{(stat.)} ^{+1.15}_{-1.05}\text{(syst.)}$ \\
\hline
\end{tabular}
\end{table}

According to Refs.~\cite{Mrowczynski:2019yrr, ant:3, Mrowczynski:2020ugu}, the source sizes of deuteron emitted through a direct hadronization process should follow $m_{\rm T}$ scaling and adhere to the scaling behavior of non-composite particles, whereas deuterons produced via the coalescence process should exhibit larger femtoscopic source sizes than expected from the scaling of individual hadrons at the corresponding $m_{\rm T}$. However, given the current statistical and systematic uncertainties, it is not possible to discern any significant differences in the scaling behavior. All measured values fall within the systematic uncertainty bands of the  $m_{\rm T}$ scaling trend for thermally produced particles. This suggests that there are no significant deviations in the femtoscopic source behavior between (anti\nobreakdash-)deuterons and other baryons. While the assumption of a common scaling appears valid for deuterons, the present statistical and systematic uncertainties prevent drawing any definitive conclusions about their production mechanisms.
The common behavior of deuterons following $m_{\rm T}$ scaling of non-composite particles was also observed in pp collisions~\cite{bhawani}. It is because under this assumption, source sizes calculated from common $m_{\rm T}$ scaling measured in elementary collisions~\cite{alice2020} enabled theoretical models to accurately describe the kaon--deuteron and proton--deuteron correlation functions.

\section{Uncertainties}
\label{sec:syst_unc}
The systematic uncertainties of the measured correlation functions were evaluated for each $k^*$ interval by varying event, track, and PID selection criteria. The correlation functions were recalculated using the default settings described in Sec.~\ref{sec:DataAnalysis} by adjusting one parameter at a time to the settings summarized in Table~\ref{tab:trackcuts2}. 
The final uncertainty at a given $k^*$ is estimated as the standard deviation of a uniform distribution defined by the range between the maximum and minimum values of the correlation function.
\begin{table}[tb!h]
\centering
\caption{Systematic variations of single track selection criteria for pions and (anti\nobreakdash-)deuterons.}
\label{tab:trackcuts2}
\begin{tabular}{l|l|l|l}
  \hline
  \hline
   Selection criterium & \multicolumn{3}{c}{Variation}\\ \hline
 
  \multicolumn{4}{c}{$\uppi^+$, $\uppi^-$, d, $\rm \overline{d}$} \\ \hline  
   $|\eta|$ & \multicolumn{3}{c}{$< 0.77; 0.83$ }\\ \hline
   $|V_{\text{Z}}|$ & \multicolumn{3}{c}{$< 8$~cm }\\ \hline   
   N space points TPC & \multicolumn{3}{c}{$\ge60; 80$ }\\ \hline
   Number of events for mixing & \multicolumn{3}{c}{$5$ }\\ \hline
   Merge fraction cut & \multicolumn{3}{c}{$3\%$ }\\ \hline
  \multicolumn{4}{c}{$\uppi^+$, $\uppi^-$} \\ \hline
    $N_{\sigma,\rm TPC}$ ($p < 0.5$~GeV/$c$) & \multicolumn{3}{c}{$< 3.5$} \\ \hline
    $\sqrt{N_{\rm \sigma, TOF}^{2} + N_{\rm \sigma, TPC}^{2}}$ 
    ($  p > 0.5$~GeV/$c$) 
    & \multicolumn{3}{c}{$< 3.5$}  \\ 
    \hline
  \multicolumn{4}{c}{d, $\rm \overline{d}$} \\ \hline
    $N_{\sigma,\rm TPC}$ ($p < 1.3$~GeV/$c$) & \multicolumn{3}{c}{$< 2.5$} \\ \hline
    $N_{\rm \sigma, TOF}$,  $N_{\rm \sigma, TPC}$  ( $p >$ 1.3~GeV/$c$) & \multicolumn{3}{c}{$< 2.5$} \\ \hline
    $\rm DCA_{\rm XY}$ to primary vertex (only for d)& \multicolumn{3}{c}{$< 0.1$~cm} \\ \hline  \hline
\end{tabular}
\end{table}

The statistical uncertainties of the $R_{\rm \uppi d}$ and $R_{\rm d}$ source are obtained by subsampling data points within the Gaussian distribution of the statistical error. Each randomly resampled correlation function is then fitted providing radii of certain variation. The procedure has been repeated one thousand times and the final statistical uncertainty corresponds to the 1$\sigma$ range of the resampled radii distribution. 

The systematic uncertainties of the $R_{\rm \uppi d}$ and $R_{\rm d}$ values obtained from the fit to the correlation functions are accounted for by incorporating the systematic uncertainties of the data points described above through the subsampling method, similar to the treatment of statistical uncertainties. Additionally, the sampled points are fitted while systematically varying the fitting settings to account for different fit ranges, non-femtoscopic background corrections, normalization ranges, and the value of the total correction factor, addressing potential underestimation or overestimation of these settings. The fitting variation settings are summarized in Table~\ref{tab:trackcuts3}. The final systematic uncertainty is derived from the 1$\sigma$ range of the radii distribution after subsampling process and using varied fitting settings.
The values of $R_{\rm d}$ also depend on the calculated value of the  femtoscopic source for pions, $R_{\rm \uppi}$. This contribution has been derived separately by repetitive fitting using the subsampling method, where the fixed value of $R_{\rm{\uppi}}$ in the fitting process is randomly varied according to a Gaussian distribution. The mean and width of the Gaussian distribution are equal to the central value of the pion’s source calculation and the uncertainties of their measurements, respectively. The uncertainty of deuterons source measurements arising from the calculation for pions corresponds to the 1$\sigma$ range of $R_d$ values obtained from such sunsampled fits.

\begin{table}[tb!h]
\centering
\caption{Systematic variations of the fitting settings.}
\label{tab:trackcuts3}
\begin{tabular}{l|l|l|l}
  \hline
  \hline
   Parameter& \multicolumn{3}{c}{Variation}\\ \hline
   Fit range & \multicolumn{3}{c}{[0-0.10]; [0-0.12]; [0--0.15]~GeV/$c$}\\ \hline
   Background parametrization  & \multicolumn{3}{c}{$1^{st}$ order polynomial, [0.15--0.35]~GeV/$c$ }\\ 
     & \multicolumn{3}{c}{$1^{st}$ order polynomial, [0.10--0.30]~GeV/$c$}\\
     & \multicolumn{3}{c}{$3^{rd}$ order polynomial, [0.10--0.90]~GeV/$c$ }\\ \hline
   Normalization range & \multicolumn{3}{c}{[0.15-0.35]; [0.10--0.30]~GeV/$c$}\\ \hline
   $corr$ factor& \multicolumn{3}{c}{[--5$\%$, +5$\%$]}\\ \hline  \hline

\end{tabular}
\end{table}

\section{\label{sec:level11} Summary}
This paper presents the first-ever study of non-identical particle femtoscopy for pion–deuteron pairs in large systems produced in Pb--Pb collisions at $\sqrt{s_{\text{NN}}}~=$~5.02 TeV. The measured pion--deuteron femtoscopic radii are equal to \mbox{10.35 $\pm$ $0.30$ (stat.) $^{+0.67}_{-0.65}$ (syst.)~fm}, \mbox{8.68 $\pm$ $0.53$ (stat.) $^{+0.48}_{-1.22}$ (syst.)~fm}, \mbox{6.95 $\pm$ $0.47$ (stat.) $^{+0.46}_{-0.58}$ (syst.)~fm} for centralities 0--10$\%$, 10--30$\%$, 30--50$\%$, respectively. By using as an input the $m_{\rm T}$ scaling observed in previous measurements on the single-pion source radii, the femtoscopic sizes of (anti\nobreakdash-)deuteron are calculated for the first time for the three measured centrality classes. The measured femtoscopic radii of (anti\nobreakdash-)deuteron are \mbox{3.88 $\pm$ $0.48$ (stat.) $^{+1.10}_{-0.64}$ (syst.)~fm}, \mbox{$3.29\pm1.21$ (stat.) $^{+0.97}_{-1.52}$ (syst.)~fm}, \mbox{2.72 $\pm$ $0.99$ (stat.) $^{+1.15}_{-1.05}$ (syst.)~fm} for centralities 0--10$\%$, 10--30$\%$, 30--50$\%$, respectively. Within statistical and systematic uncertainties, deuterons and antideuterons follow the $m_{\rm T}$ scaling trend, similar to other hadrons such as pions, kaons, and protons. This provides experimental confirmation of the validity of extending $m_{\rm T}$ scaling to these light ions. Therefore, it opens new possibilities for future interaction studies involving deuterons, enabling a more precise determination of particle-emitting source properties using other abundant particle species. 
Nevertheless, the current precision of the measured femtoscopic sources of deuterons does not allow for definitive conclusions regarding their production mechanism based on their deviation from or alignment with $m_{\rm T}$ scaling, which would indicate either thermal or coalescence production. 

Moreover, this study demonstrates that by employing non-identical particle femtoscopy with highly abundant particles produced in collisions, such as pions, we can investigate the properties of the source of rarer particle species. An identical particle femtoscopy analysis of these species would pose a major challenge due to the low abundances and large statistical uncertainties involved.   
 This measurement opens the possibility to study rare sources, such as multi-strange baryons or even D-mesons, with the substantially increased data set collected by the upgraded ALICE detector in Run 3. 



\newenvironment{acknowledgement}{\relax}{\relax}
\begin{acknowledgement}
\section*{Acknowledgements}

The ALICE Collaboration would like to thank all its engineers and technicians for their invaluable contributions to the construction of the experiment and the CERN accelerator teams for the outstanding performance of the LHC complex.
The ALICE Collaboration gratefully acknowledges the resources and support provided by all Grid centres and the Worldwide LHC Computing Grid (WLCG) collaboration.
The ALICE Collaboration acknowledges the following funding agencies for their support in building and running the ALICE detector:
A. I. Alikhanyan National Science Laboratory (Yerevan Physics Institute) Foundation (ANSL), State Committee of Science and World Federation of Scientists (WFS), Armenia;
Austrian Academy of Sciences, Austrian Science Fund (FWF): [M 2467-N36] and Nationalstiftung f\"{u}r Forschung, Technologie und Entwicklung, Austria;
Ministry of Communications and High Technologies, National Nuclear Research Center, Azerbaijan;
Conselho Nacional de Desenvolvimento Cient\'{\i}fico e Tecnol\'{o}gico (CNPq), Financiadora de Estudos e Projetos (Finep), Funda\c{c}\~{a}o de Amparo \`{a} Pesquisa do Estado de S\~{a}o Paulo (FAPESP) and Universidade Federal do Rio Grande do Sul (UFRGS), Brazil;
Bulgarian Ministry of Education and Science, within the National Roadmap for Research Infrastructures 2020-2027 (object CERN), Bulgaria;
Ministry of Education of China (MOEC) , Ministry of Science \& Technology of China (MSTC) and National Natural Science Foundation of China (NSFC), China;
Ministry of Science and Education and Croatian Science Foundation, Croatia;
Centro de Aplicaciones Tecnol\'{o}gicas y Desarrollo Nuclear (CEADEN), Cubaenerg\'{\i}a, Cuba;
Ministry of Education, Youth and Sports of the Czech Republic, Czech Republic;
The Danish Council for Independent Research | Natural Sciences, the VILLUM FONDEN and Danish National Research Foundation (DNRF), Denmark;
Helsinki Institute of Physics (HIP), Finland;
Commissariat \`{a} l'Energie Atomique (CEA) and Institut National de Physique Nucl\'{e}aire et de Physique des Particules (IN2P3) and Centre National de la Recherche Scientifique (CNRS), France;
Bundesministerium f\"{u}r Bildung und Forschung (BMBF) and GSI Helmholtzzentrum f\"{u}r Schwerionenforschung GmbH, Germany;
General Secretariat for Research and Technology, Ministry of Education, Research and Religions, Greece;
National Research, Development and Innovation Office, Hungary;
Department of Atomic Energy Government of India (DAE), Department of Science and Technology, Government of India (DST), University Grants Commission, Government of India (UGC) and Council of Scientific and Industrial Research (CSIR), India;
National Research and Innovation Agency - BRIN, Indonesia;
Istituto Nazionale di Fisica Nucleare (INFN), Italy;
Japanese Ministry of Education, Culture, Sports, Science and Technology (MEXT) and Japan Society for the Promotion of Science (JSPS) KAKENHI, Japan;
Consejo Nacional de Ciencia (CONACYT) y Tecnolog\'{i}a, through Fondo de Cooperaci\'{o}n Internacional en Ciencia y Tecnolog\'{i}a (FONCICYT) and Direcci\'{o}n General de Asuntos del Personal Academico (DGAPA), Mexico;
Nederlandse Organisatie voor Wetenschappelijk Onderzoek (NWO), Netherlands;
The Research Council of Norway, Norway;
Pontificia Universidad Cat\'{o}lica del Per\'{u}, Peru;
Ministry of Science and Higher Education, National Science Centre and WUT ID-UB, Poland;
Korea Institute of Science and Technology Information and National Research Foundation of Korea (NRF), Republic of Korea;
Ministry of Education and Scientific Research, Institute of Atomic Physics, Ministry of Research and Innovation and Institute of Atomic Physics and Universitatea Nationala de Stiinta si Tehnologie Politehnica Bucuresti, Romania;
Ministerstvo skolstva, vyskumu, vyvoja a mladeze SR, Slovakia;
National Research Foundation of South Africa, South Africa;
Swedish Research Council (VR) and Knut \& Alice Wallenberg Foundation (KAW), Sweden;
European Organization for Nuclear Research, Switzerland;
Suranaree University of Technology (SUT), National Science and Technology Development Agency (NSTDA) and National Science, Research and Innovation Fund (NSRF via PMU-B B05F650021), Thailand;
Turkish Energy, Nuclear and Mineral Research Agency (TENMAK), Turkey;
National Academy of  Sciences of Ukraine, Ukraine;
Science and Technology Facilities Council (STFC), United Kingdom;
National Science Foundation of the United States of America (NSF) and United States Department of Energy, Office of Nuclear Physics (DOE NP), United States of America.
In addition, individual groups or members have received support from:
Czech Science Foundation (grant no. 23-07499S), Czech Republic;
FORTE project, reg.\ no.\ CZ.02.01.01/00/22\_008/0004632, Czech Republic, co-funded by the European Union, Czech Republic;
European Research Council (grant no. 950692), European Union;
Deutsche Forschungs Gemeinschaft (DFG, German Research Foundation) ``Neutrinos and Dark Matter in Astro- and Particle Physics'' (grant no. SFB 1258), Germany;
ICSC - National Research Center for High Performance Computing, Big Data and Quantum Computing and FAIR - Future Artificial Intelligence Research, funded by the NextGenerationEU program (Italy).

\end{acknowledgement}

\bibliographystyle{utphys}   
\bibliography{apssamp}

\newpage
\appendix
\section{Lednick\'y--Lyuboshitz formalism}
\label{sec:appendix:LL}
The pair wave function, denoted as $\Psi(\vec{k}^{},\vec{r}^{})$ in Eq.~(\ref{eq:cf_def}), depends on the two-particle interaction. Specifically, in the context of charged pions interacting with deuterons, both the Coulomb and strong forces come into play. In this scenario, the interaction between two non-identical particles is described by the Bethe--Salpeter amplitude, which corresponds to the solution of the quantum scattering problem approached in reverse time (indicated by the ``$-$'' sign preceding $k^*$)
\vspace{0.25cm}\\
\begin{equation}
\Psi^{(+)}_{-{\vec{k}^{*}}}({\vec{{r}^{*}}}) = \sqrt{A_{\rm C} (\varepsilon)}
\frac{1}{\sqrt{2}} \left [ {\rm e}^{-{\rm i} \vec{ k}^{*}\cdot{\vec{r}^{*}}} {\rm F}(-{\rm i} \varepsilon, 1,
  {\rm i} \zeta^{+}) + f_{\rm C}(\vec{k}^*)\frac{\tilde{G}(\rho,\varepsilon)}{{r}^*} \right ],
\label{eq:fullpsi2}
\end{equation}
\vspace{0.01cm}\\
where $A_{\rm C}$ is the Gamow factor,
$\varepsilon = 1/(k^{*} a_{\rm C})$, 
$\zeta^{\pm} = k^{*} r^{*} (1 \pm \cos{\theta^{*}})$, 
$\rm F$ is the confluent hypergeometric function, and $\tilde{G}$ is the combination of the regular and singular s-wave Coulomb functions.
The angle $\theta^{*}$  delineates the relationship between the pair's relative momentum ($k^*$) and relative position within the pair's rest frame, whereas $a_{\rm C}$ represents the Bohr radius of the pair. The term $f_{\rm C}(k^*)$ is  the scattering amplitude resulting from strong interaction, adjusted to accommodate the Coulomb component

\begin{equation}
\label{eq:scattering_amplitude}
    f^{-1}_C(k^*) = \frac{1}{f_0} + \frac{1}{2}d_0{k^*}^2-\frac{2}{a_C}h(k^* a_C) -\text{i}k^* a_C,
\end{equation}
where $h(\varepsilon)=\varepsilon^2\sum\limits_{n=1}^{ \infty}[n(n^2+\varepsilon^2)]^{-1}
-\mathrm{{\gamma}} -\ln|\varepsilon|$
($\mathrm{{\gamma}} = 0.5772$ represents the Euler constant).

The correlation function is computed by numerical integration of the source function $S(r^*)$, which is modeled by a three-dimensional Gaussian within the PRF ($\vec{p_1}^*=-\vec{p_2}^*$). This integration is performed in conjunction with the utilization of the Bethe--Salpeter amplitude provided in Eq.~(\ref{eq:fullpsi2}) and with the Coulomb-modified scattering amplitude defined in Eq.~(\ref{eq:scattering_amplitude})~\cite{Kisiel:2009eh,Kisiel:2018wie}.


\newpage
\section{The ALICE Collaboration}
\label{app:collab}
\begin{flushleft} 
\small

S.~Acharya\,\orcidlink{0000-0002-9213-5329}\,$^{\rm 50}$, 
A.~Agarwal$^{\rm 133}$, 
G.~Aglieri Rinella\,\orcidlink{0000-0002-9611-3696}\,$^{\rm 32}$, 
L.~Aglietta\,\orcidlink{0009-0003-0763-6802}\,$^{\rm 24}$, 
M.~Agnello\,\orcidlink{0000-0002-0760-5075}\,$^{\rm 29}$, 
N.~Agrawal\,\orcidlink{0000-0003-0348-9836}\,$^{\rm 25}$, 
Z.~Ahammed\,\orcidlink{0000-0001-5241-7412}\,$^{\rm 133}$, 
S.~Ahmad\,\orcidlink{0000-0003-0497-5705}\,$^{\rm 15}$, 
S.U.~Ahn\,\orcidlink{0000-0001-8847-489X}\,$^{\rm 71}$, 
I.~Ahuja\,\orcidlink{0000-0002-4417-1392}\,$^{\rm 36}$
% ...

\section*{Affiliation Notes}

$^{\rm I}$ Deceased\\
$^{\rm II}$ Also at: Max-Planck-Institut fur Physik, Munich, Germany\\
$^{\rm III}$ Also at: Italian National Agency for New Technologies, Energy and Sustainable Economic Development (ENEA), Bologna, Italy\\
$^{\rm IV}$ Also at: Instituto de Fisica da Universidade de Sao Paulo\\
$^{\rm V}$ Also at: Dipartimento DET del Politecnico di Torino, Turin, Italy\\
$^{\rm VI}$ Also at: Department of Applied Physics, Aligarh Muslim University, Aligarh, India\\
$^{\rm VII}$ Also at: Institute of Theoretical Physics, University of Wroclaw, Poland\\
$^{\rm VIII}$ Also at: Facultad de Ciencias, Universidad Nacional Aut\'{o}noma de M\'{e}xico, Mexico City, Mexico\\

\section*{Collaboration Institutes}

$^{1}$ A.I. Alikhanyan National Science Laboratory (Yerevan Physics Institute) Foundation, Yerevan, Armenia\\
$^{2}$ AGH University of Krakow, Cracow, Poland\\
$^{3}$ Bogolyubov Institute for Theoretical Physics, National Academy of Sciences of Ukraine, Kiev, Ukraine\\
$^{4}$ Bose Institute, Department of Physics  and Centre for Astroparticle Physics and Space Science (CAPSS), Kolkata, India\\
$^{5}$ California Polytechnic State University, San Luis Obispo, California, United States\\
$^{6}$ Central China Normal University, Wuhan, China\\
$^{7}$ Centro de Aplicaciones Tecnol\'{o}gicas y Desarrollo Nuclear (CEADEN), Havana, Cuba\\
$^{8}$ Centro de Investigaci\'{o}n y de Estudios Avanzados (CINVESTAV), Mexico City and M\'{e}rida, Mexico\\
$^{9}$ Chicago State University, Chicago, Illinois, United States\\
$^{10}$ China Nuclear Data Center, China Institute of Atomic Energy, Beijing, China\\
$^{11}$ China University of Geosciences, Wuhan, China\\
$^{12}$ Chungbuk National University, Cheongju, Republic of Korea\\
$^{13}$ Comenius University Bratislava, Faculty of Mathematics, Physics and Informatics, Bratislava, Slovak Republic\\
$^{14}$ Creighton University, Omaha, Nebraska, United States\\
$^{15}$ Department of Physics, Aligarh Muslim University, Aligarh, India\\
$^{16}$ Department of Physics, Pusan National University, Pusan, Republic of Korea\\
$^{17}$ Department of Physics, Sejong University, Seoul, Republic of Korea\\
$^{18}$ Department of Physics, University of California, Berkeley, California, United States\\
$^{19}$ Department of Physics, University of Oslo, Oslo, Norway\\
$^{20}$ Department of Physics and Technology, University of Bergen, Bergen, Norway\\
$^{21}$ Dipartimento di Fisica, Universit\`{a} di Pavia, Pavia, Italy\\
$^{22}$ Dipartimento di Fisica dell'Universit\`{a} and Sezione INFN, Cagliari, Italy\\
$^{23}$ Dipartimento di Fisica dell'Universit\`{a} and Sezione INFN, Trieste, Italy\\
$^{24}$ Dipartimento di Fisica dell'Universit\`{a} and Sezione INFN, Turin, Italy\\
$^{25}$ Dipartimento di Fisica e Astronomia dell'Universit\`{a} and Sezione INFN, Bologna, Italy\\
$^{26}$ Dipartimento di Fisica e Astronomia dell'Universit\`{a} and Sezione INFN, Catania, Italy\\
$^{27}$ Dipartimento di Fisica e Astronomia dell'Universit\`{a} and Sezione INFN, Padova, Italy\\
$^{28}$ Dipartimento di Fisica `E.R.~Caianiello' dell'Universit\`{a} and Gruppo Collegato INFN, Salerno, Italy\\
$^{29}$ Dipartimento DISAT del Politecnico and Sezione INFN, Turin, Italy\\
$^{30}$ Dipartimento di Scienze MIFT, Universit\`{a} di Messina, Messina, Italy\\
$^{31}$ Dipartimento Interateneo di Fisica `M.~Merlin' and Sezione INFN, Bari, Italy\\
$^{32}$ European Organization for Nuclear Research (CERN), Geneva, Switzerland\\
$^{33}$ Faculty of Electrical Engineering, Mechanical Engineering and Naval Architecture, University of Split, Split, Croatia\\
$^{34}$ Faculty of Nuclear Sciences and Physical Engineering, Czech Technical University in Prague, Prague, Czech Republic\\
$^{35}$ Faculty of Physics, Sofia University, Sofia, Bulgaria\\
$^{36}$ Faculty of Science, P.J.~\v{S}af\'{a}rik University, Ko\v{s}ice, Slovak Republic\\
$^{37}$ Faculty of Technology, Environmental and Social Sciences, Bergen, Norway\\
$^{38}$ Frankfurt Institute for Advanced Studies, Johann Wolfgang Goethe-Universit\"{a}t Frankfurt, Frankfurt, Germany\\
$^{39}$ Fudan University, Shanghai, China\\
$^{40}$ Gangneung-Wonju National University, Gangneung, Republic of Korea\\
$^{41}$ Gauhati University, Department of Physics, Guwahati, India\\
$^{42}$ Helmholtz-Institut f\"{u}r Strahlen- und Kernphysik, Rheinische Friedrich-Wilhelms-Universit\"{a}t Bonn, Bonn, Germany\\
$^{43}$ Helsinki Institute of Physics (HIP), Helsinki, Finland\\
$^{44}$ High Energy Physics Group,  Universidad Aut\'{o}noma de Puebla, Puebla, Mexico\\
$^{45}$ Horia Hulubei National Institute of Physics and Nuclear Engineering, Bucharest, Romania\\
$^{46}$ HUN-REN Wigner Research Centre for Physics, Budapest, Hungary\\
$^{47}$ Indian Institute of Technology Bombay (IIT), Mumbai, India\\
$^{48}$ Indian Institute of Technology Indore, Indore, India\\
$^{49}$ INFN, Laboratori Nazionali di Frascati, Frascati, Italy\\
$^{50}$ INFN, Sezione di Bari, Bari, Italy\\
$^{51}$ INFN, Sezione di Bologna, Bologna, Italy\\
$^{52}$ INFN, Sezione di Cagliari, Cagliari, Italy\\
$^{53}$ INFN, Sezione di Catania, Catania, Italy\\
$^{54}$ INFN, Sezione di Padova, Padova, Italy\\
$^{55}$ INFN, Sezione di Pavia, Pavia, Italy\\
$^{56}$ INFN, Sezione di Torino, Turin, Italy\\
$^{57}$ INFN, Sezione di Trieste, Trieste, Italy\\
$^{58}$ Inha University, Incheon, Republic of Korea\\
$^{59}$ Institute for Gravitational and Subatomic Physics (GRASP), Utrecht University/Nikhef, Utrecht, Netherlands\\
$^{60}$ Institute of Experimental Physics, Slovak Academy of Sciences, Ko\v{s}ice, Slovak Republic\\
$^{61}$ Institute of Physics, Homi Bhabha National Institute, Bhubaneswar, India\\
$^{62}$ Institute of Physics of the Czech Academy of Sciences, Prague, Czech Republic\\
$^{63}$ Institute of Space Science (ISS), Bucharest, Romania\\
$^{64}$ Institut f\"{u}r Kernphysik, Johann Wolfgang Goethe-Universit\"{a}t Frankfurt, Frankfurt, Germany\\
$^{65}$ Instituto de Ciencias Nucleares, Universidad Nacional Aut\'{o}noma de M\'{e}xico, Mexico City, Mexico\\
$^{66}$ Instituto de F\'{i}sica, Universidade Federal do Rio Grande do Sul (UFRGS), Porto Alegre, Brazil\\
$^{67}$ Instituto de F\'{\i}sica, Universidad Nacional Aut\'{o}noma de M\'{e}xico, Mexico City, Mexico\\
$^{68}$ iThemba LABS, National Research Foundation, Somerset West, South Africa\\
$^{69}$ Jeonbuk National University, Jeonju, Republic of Korea\\
$^{70}$ Johann-Wolfgang-Goethe Universit\"{a}t Frankfurt Institut f\"{u}r Informatik, Fachbereich Informatik und Mathematik, Frankfurt, Germany\\
$^{71}$ Korea Institute of Science and Technology Information, Daejeon, Republic of Korea\\
$^{72}$ Laboratoire de Physique Subatomique et de Cosmologie, Universit\'{e} Grenoble-Alpes, CNRS-IN2P3, Grenoble, France\\
$^{73}$ Lawrence Berkeley National Laboratory, Berkeley, California, United States\\
$^{74}$ Lund University Department of Physics, Division of Particle Physics, Lund, Sweden\\
$^{75}$ Nagasaki Institute of Applied Science, Nagasaki, Japan\\
$^{76}$ Nara Women{'}s University (NWU), Nara, Japan\\
$^{77}$ National and Kapodistrian University of Athens, School of Science, Department of Physics , Athens, Greece\\
$^{78}$ National Centre for Nuclear Research, Warsaw, Poland\\
$^{79}$ National Institute of Science Education and Research, Homi Bhabha National Institute, Jatni, India\\
$^{80}$ National Nuclear Research Center, Baku, Azerbaijan\\
$^{81}$ National Research and Innovation Agency - BRIN, Jakarta, Indonesia\\
$^{82}$ Niels Bohr Institute, University of Copenhagen, Copenhagen, Denmark\\
$^{83}$ Nikhef, National institute for subatomic physics, Amsterdam, Netherlands\\
$^{84}$ Nuclear Physics Group, STFC Daresbury Laboratory, Daresbury, United Kingdom\\
$^{85}$ Nuclear Physics Institute of the Czech Academy of Sciences, Husinec-\v{R}e\v{z}, Czech Republic\\
$^{86}$ Oak Ridge National Laboratory, Oak Ridge, Tennessee, United States\\
$^{87}$ Ohio State University, Columbus, Ohio, United States\\
$^{88}$ Physics department, Faculty of science, University of Zagreb, Zagreb, Croatia\\
$^{89}$ Physics Department, Panjab University, Chandigarh, India\\
$^{90}$ Physics Department, University of Jammu, Jammu, India\\
$^{91}$ Physics Program and International Institute for Sustainability with Knotted Chiral Meta Matter (WPI-SKCM$^{2}$), Hiroshima University, Hiroshima, Japan\\
$^{92}$ Physikalisches Institut, Eberhard-Karls-Universit\"{a}t T\"{u}bingen, T\"{u}bingen, Germany\\
$^{93}$ Physikalisches Institut, Ruprecht-Karls-Universit\"{a}t Heidelberg, Heidelberg, Germany\\
$^{94}$ Physik Department, Technische Universit\"{a}t M\"{u}nchen, Munich, Germany\\
$^{95}$ Politecnico di Bari and Sezione INFN, Bari, Italy\\
$^{96}$ Research Division and ExtreMe Matter Institute EMMI, GSI Helmholtzzentrum f\"ur Schwerionenforschung GmbH, Darmstadt, Germany\\
$^{97}$ Saga University, Saga, Japan\\
$^{98}$ Saha Institute of Nuclear Physics, Homi Bhabha National Institute, Kolkata, India\\
$^{99}$ School of Physics and Astronomy, University of Birmingham, Birmingham, United Kingdom\\
$^{100}$ Secci\'{o}n F\'{\i}sica, Departamento de Ciencias, Pontificia Universidad Cat\'{o}lica del Per\'{u}, Lima, Peru\\
$^{101}$ Stefan Meyer Institut f\"{u}r Subatomare Physik (SMI), Vienna, Austria\\
$^{102}$ SUBATECH, IMT Atlantique, Nantes Universit\'{e}, CNRS-IN2P3, Nantes, France\\
$^{103}$ Sungkyunkwan University, Suwon City, Republic of Korea\\
$^{104}$ Suranaree University of Technology, Nakhon Ratchasima, Thailand\\
$^{105}$ Technical University of Ko\v{s}ice, Ko\v{s}ice, Slovak Republic\\
$^{106}$ The Henryk Niewodniczanski Institute of Nuclear Physics, Polish Academy of Sciences, Cracow, Poland\\
$^{107}$ The University of Texas at Austin, Austin, Texas, United States\\
$^{108}$ Universidad Aut\'{o}noma de Sinaloa, Culiac\'{a}n, Mexico\\
$^{109}$ Universidade de S\~{a}o Paulo (USP), S\~{a}o Paulo, Brazil\\
$^{110}$ Universidade Estadual de Campinas (UNICAMP), Campinas, Brazil\\
$^{111}$ Universidade Federal do ABC, Santo Andre, Brazil\\
$^{112}$ Universitatea Nationala de Stiinta si Tehnologie Politehnica Bucuresti, Bucharest, Romania\\
$^{113}$ University of Derby, Derby, United Kingdom\\
$^{114}$ University of Houston, Houston, Texas, United States\\
$^{115}$ University of Jyv\"{a}skyl\"{a}, Jyv\"{a}skyl\"{a}, Finland\\
$^{116}$ University of Kansas, Lawrence, Kansas, United States\\
$^{117}$ University of Liverpool, Liverpool, United Kingdom\\
$^{118}$ University of Science and Technology of China, Hefei, China\\
$^{119}$ University of South-Eastern Norway, Kongsberg, Norway\\
$^{120}$ University of Tennessee, Knoxville, Tennessee, United States\\
$^{121}$ University of the Witwatersrand, Johannesburg, South Africa\\
$^{122}$ University of Tokyo, Tokyo, Japan\\
$^{123}$ University of Tsukuba, Tsukuba, Japan\\
$^{124}$ Universit\"{a}t M\"{u}nster, Institut f\"{u}r Kernphysik, M\"{u}nster, Germany\\
$^{125}$ Universit\'{e} Clermont Auvergne, CNRS/IN2P3, LPC, Clermont-Ferrand, France\\
$^{126}$ Universit\'{e} de Lyon, CNRS/IN2P3, Institut de Physique des 2 Infinis de Lyon, Lyon, France\\
$^{127}$ Universit\'{e} de Strasbourg, CNRS, IPHC UMR 7178, F-67000 Strasbourg, France, Strasbourg, France\\
$^{128}$ Universit\'{e} Paris-Saclay, Centre d'Etudes de Saclay (CEA), IRFU, D\'{e}partment de Physique Nucl\'{e}aire (DPhN), Saclay, France\\
$^{129}$ Universit\'{e}  Paris-Saclay, CNRS/IN2P3, IJCLab, Orsay, France\\
$^{130}$ Universit\`{a} degli Studi di Foggia, Foggia, Italy\\
$^{131}$ Universit\`{a} del Piemonte Orientale, Vercelli, Italy\\
$^{132}$ Universit\`{a} di Brescia, Brescia, Italy\\
$^{133}$ Variable Energy Cyclotron Centre, Homi Bhabha National Institute, Kolkata, India\\
$^{134}$ Warsaw University of Technology, Warsaw, Poland\\
$^{135}$ Wayne State University, Detroit, Michigan, United States\\
$^{136}$ Yale University, New Haven, Connecticut, United States\\
$^{137}$ Yildiz Technical University, Istanbul, Turkey\\
$^{138}$ Yonsei University, Seoul, Republic of Korea\\
$^{139}$ Affiliated with an institute formerly covered by a cooperation agreement with CERN\\
$^{140}$ Affiliated with an international laboratory covered by a cooperation agreement with CERN.\\

\end{flushleft}

\end{document}